\documentclass[showpacs,prl,aps,superscriptaddress,prl,twocolumn,preprintnumbers,letterpaper]{revtex4}
\usepackage{amsmath,amssymb,bm}
\usepackage{epsfig}
\usepackage{graphicx}
\usepackage{amsfonts}
\usepackage{epstopdf}
\usepackage{color}

\renewcommand{\part}{{\rm part}}

\renewcommand{\vec}{\boldsymbol}

\begin{document}

\title{Chiral vortical wave and induced flavor charge transport \\
in a rotating quark-gluon plasma}

 \author{Yin Jiang}
\email{jiangyin@indiana.edu}
\affiliation{ Physics Department and Center for Exploration of Energy and Matter,
Indiana University, 2401 N Milo B. Sampson Lane, Bloomington, IN 47408, USA.}

\author{Xu-Guang Huang}
\email{huangxuguang@fudan.edu.cn}
\affiliation{Physics Department and Center for Particle Physics and Field Theory, Fudan University, Shanghai 200433, China.}

\author{Jinfeng Liao}
\email{liaoji@indiana.edu}
\affiliation{ Physics Department and Center for Exploration of Energy and Matter,
Indiana University, 2401 N Milo B. Sampson Lane, Bloomington, IN 47408, USA.}
\affiliation{RIKEN BNL Research Center, Bldg. 510A, Brookhaven National Laboratory, Upton, NY 11973, USA.}

\date{\today}

\begin{abstract}
We show the existence of a new gapless collective excitation in a rotating fluid system with chiral fermions, named as the Chiral Vortical Wave (CVW). The CVW has its microscopic origin at the quantum anomaly and macroscopically arises from interplay between vector and axial charge fluctuations induced by vortical effects.  The wave equation is obtained both from hydrodynamic current equations and from chiral kinetic theory and its solutions show nontrivial CVW-induced charge transport from different initial conditions. Using the rotating quark-gluon plasma in heavy ion collisions as a concrete example, we show the formation of induced flavor quadrupole in QGP and estimate the elliptic flow splitting effect for $\Lambda$ baryons that may be experimentally measured.
\end{abstract}
\maketitle

{\it Introduction.---} Anomalous transport effects in many body systems with chiral fermions have generated great interests recently. Such phenomena span a wide range of physical systems~\cite{Kharzeev:2015kna,Miransky:2015ava}, from semimetals to cold atomic systems, and from hot quark-gluon plasma (QGP) created in heavy ion collisions to cold dense matter in neutron stars. These systems provide possible environments with nonzero macroscopic chirality and can manifest the microscopic chiral anomaly in macroscopic transport processes that would be normally forbidden by symmetries like parity invariance.

One way to induce the anomalous transport effects in such chiral system is to apply external electromagnetic fields. A famous example is the so-called Chiral Magnetic Effect (CME) in which an electric current can be generated in parallel to an external magnetic field . The CME could lead to experimentally measurable effects both for the QGP in heavy ion collisions~\cite{Kharzeev:2004ey,Kharzeev:2007tn,Kharzeev:2007jp,STAR:2009uh,Adamczyk:2014mzf,Adamczyk:2013hsi,Ajitanand:2010rc,Abelev:2012pa,Wang:2009kd,Bzdak:2009fc,Pratt,Muller,Deng:2014uja} and for certain Dirac and Weyl semimetals~\cite{Kharzeev:metal}. Other interesting examples include e.g. the Chiral Separation Effect (CSE)~\cite{son:2004tq,Metlitski:2005pr}, and the Chiral Electric Separation Effect (CESE)~\cite{Huang:2013iia,Jiang:2014ura}. For reviews see e.g.~\cite{Kharzeev:2013ffa,Liao:2014ava,Bzdak:2012ia}.

The anomalous transport effects can also occur when the fluid undergoes a global rotation quantified by a vorticity $\vec \omega =\frac{1}{2}\vec\nabla \times \vec v$ where $\vec v$ is the flow velocity field. Such vortical effects were suggested in \cite{Kharzeev:2007tn} and quantified in holographic models~\cite{Banerjee:2008th,Erdmenger:2008rm,Torabian:2009qk} and the anomalous hydrodynamic framework~\cite{Son:2009tf}. The so-called Chiral Vortical Effect (CVE) quantifies the generated vector current ${\vec J}_V$ as~\cite{Banerjee:2008th,Erdmenger:2008rm,Torabian:2009qk,Son:2009tf}
\begin{eqnarray} \label{eq_VV}
&&\vec{J}_V=\frac{1}{\pi^2}\mu\mu_5\vec{\omega} \,\, ,
\end{eqnarray}
  and the generated axial current ${\vec J}_A$ as~\cite{Banerjee:2008th,Erdmenger:2008rm,Torabian:2009qk,Son:2009tf}
 \begin{eqnarray} \label{eq_VA}
&&\vec{J}_A=\left [ \frac{1}{6}T^2+\frac{1}{2\pi^2}(\mu^2+\mu^2_5) \right ] \vec{\omega} \,\, ,
\end{eqnarray}
where $T$ is temperature and $\mu$ and $\mu_5$ are vector and axial chemical potentials. It was suggested that such CVE may lead to baryon charge separation in heavy ion collisions~\cite{Kharzeev:2010gr,STAR2014}.

The vorticity-driven anomalous transport effects in Eqs.~(\ref{eq_VV}) and (\ref{eq_VA}) couple together the vector and axial densities/currents. Similar situation also happens in external magnetic field where the interplay between CME and CSE leads to a gapless collective mode called {\it ``Chiral Magnetic Wave'' (CMW)}~\cite{Kharzeev:2010gd,Burnier:2011bf,Adamczyk:2015eqo,Gorbar:2011ya}. In this Letter, we show that the vortical effects also induce a new wave mode for vector and axial density fluctuations which we call a {\it ``Chiral Vortical Wave'' (CVW)}. We will derive this new wave equation and determine the CVW propagation speed in both the hydrodynamic and kinetic theory frameworks. We then show that the CVW can induce a fermion charge quadrupole distribution from initial vector density. Finally we will make predictions for possible implications of CVW in heavy ion collisions.

While we will use rotating QGP as a concrete example, the theoretical contents of CVW is in a general fashion and the proposed phenomenon is universal, being able to be realized in various systems with effective chiral fermions. We note that various other chiral effects have been explored in systems from compact stars~\cite{Gorbar:2011ya,Kaminski:2014jda,Dvornikov:2014uza,Sigl:2015xva}, to Weyl and Dirac semimetals~\cite{Gorbar:2013dha,Gorbar:2013qsa,Kharzeev:2012dc,
Basar:2013iaa,Kharzeev:2014sba}, and to spin-orbit coupled cold Fermi gases~\cite{Wang:2012,Cheuk:2012,cold_atom} (as reviewed in~\cite{Kharzeev:2015kna,Miransky:2015ava}). It would be feasible and of great interest to realize CVW as a new and independent way of manifesting chiral effects in those systems.

{\it The Chiral Vortical Wave.---} Let us start by rewriting the vortical effects (\ref{eq_VV}) and (\ref{eq_VA}) in terms of currents $\vec{J}_{L/R}=\frac{1}{2}( \vec{J}_V\mp\vec{J}_A )$:
\begin{eqnarray} \label{eq_LR}
&&\vec{J}_{L/R}=\mp \left( \frac{1}{12}T^2 + \frac{1}{4\pi^2}\mu^2_{L/R} \right) \vec{\omega} \,\, ,
\end{eqnarray}
where $\mu_{L/R}=\mu\mp\mu_5$. Intuitively the above vortical effects can be understood as follows. In the presence of global rotation, the underlying fermions experience an effective interaction of the form $\sim - \vec \omega \cdot \vec S$ in their local rest frame, with $\vec S$ the spin of fermions. This causes a {\it charge-blind} spin polarization effect (as indeed found in other context \cite{Liang:2004ph,Becattini:2007sr}), namely the fermions will have their spins preferably aligned with $\vec \omega$. 
As a result the right- or left-handed fermions will preferably have their momenta parallel or anti-parallel to $\vec \omega$, giving rise to the currents in (\ref{eq_LR}).

Let us then consider the small fluctuations of left- and right-handed densities on top of a uniform equilibrium background. For simplicity we consider $\omega$ to be constant and neglect fluctuations of temperature (which is controlled by linearized hydro equations for energy momentum tensor).
By combining the continuity equations $\partial_t n_{L/R}+\vec\nabla\cdot\vec{J}_{L/R}=0$ (here $n_{L/R} =\frac{1}{2}(J^0_V \mp J^0_A)$) with Eqs.~(\ref{eq_LR}), one obtains,
\begin{eqnarray}
\partial_t n_{L/R}= \pm\frac{1}{4\pi^2}\omega\partial_x(\mu^2_{L/R})=   \pm\frac{\omega \mu_{L/R}}{2\pi^2}\partial_x\mu_{L/R},
\end{eqnarray}
where we have set vorticity along $x$-direction $\vec \omega = \omega \hat{x}$ with $\hat{x}\equiv  \vec x/|\vec x|$. Clearly there are two modes, one for right-handed density and the other for left-handed density, that propagate in opposite direction.  For later convenience we introduce the susceptibilities for the corresponding densities: $\chi_{\mu}^{L/R} \equiv \partial n_{L/R} / \partial \mu_{L/R}$.

At this point, two possibilities may occur. For simplicity we focus on the right-handed mode below.\\
(1)  The background fluid is charge neutral, i.e. $\mu_0=0$. In this case the density fluctuation $\delta n=\chi_{0} \delta \mu$ and is governed by the nonlinear wave equation:
\begin{eqnarray}
\partial_t (\delta n) + \frac{\omega}{4\pi^2 \chi_0^2} \partial_x (\delta n^2) = 0 \,\, .
\end{eqnarray}
 This takes the form of inviscid Burgers' equation~\cite{Basar:2013iaa} whose (implicit) solution can be formally written as:
 \begin{eqnarray}
 \delta n(x,t) = F_i \left(x-\frac{\omega t \, \delta n}{2\pi^2 \chi_0^2}\right),
 \end{eqnarray}
 with $F_i(x)=\delta n(x,t=0)$ given by the initial density fluctuation. \\
 (2) The background has nonzero charge density $\mu_0\neq 0$. In this case one can linearize the equation for the evolution of small density fluctuations on top of the background density and obtain a linear wave equation:
 \begin{eqnarray} \label{eq_wave_1}
 \partial_t (\delta n) + \frac{\mu_0 \omega}{2\pi^2 \chi_{\mu_0}} \partial_x (\delta n) = 0.
 \end{eqnarray}
 This is just a usual wave equation describing a propagating mode with a gapless dispersion relation:
 \begin{eqnarray} \label{eq_cvw}
 \nu = V_{\Omega} \, |\vec k| \quad, \quad  V_{\Omega}=\frac{\mu_0 \, \omega}{2\pi^2\chi_{\mu_0}},
 \end{eqnarray}
 where $\nu$ is the wave frequency and $\vec k = k\hat{x}$ is the wave vector.
 This is the {\it Chiral Vortical Wave (CVW)} with the wave speed $V_{\Omega}$ defined above. More precisely this is the right-handed wave mode that propagates along the $\vec \omega$ direction. The left-handed wave mode propagates in opposite direction to $\vec \omega$, with a speed given by a similar formula albeit replacing $\mu_0$ and $\chi_{\mu_0}$ with the left-handed quantities.

In short, the CVW found above is essentially a hydrodynamic density wave arising from slowly varying vector and axial density fluctuations that are coupled together through vortical effects.  Possible diffusion effects can be also included by adding to the LHS of (\ref{eq_wave_1}) terms like $-(D_L\partial^2_x + D_T \partial^2_T ) (\delta n)$ where $D_{L}$ and $D_T$ are the longitudinal and transverse diffusion constants.
If the initial condition of density fluctuations is very ``lumpy'' then such diffusion effects must be taken into account.

{\it CVW from Chiral Kinetic Theory.---} Recently, the physics of chiral anomaly has been incorporated into kinetic theory framework and the anomalous transport effects such as the CME and CVE~\cite{Son:2012wh,Son:2012zy,Stephanov:2012ki,Chen:2014cla,Stephanov:2014dma,Gao:2012ix} as well as the CMW~\cite{Stephanov:2014dma} were understood in a transparent way in such chiral kinetic theory. It is therefore desirable to understand how the newly found CVW may arise in the chiral kinetic theory framework.

Let us consider a rotating system of noninteracting right-handed (denoted by ``+'') Weyl fermions as well as their left-handed (denoted by ``-'') anti-particles. (The discussion for a system of left-handed fermions with their right-handed anti-fermions will be similar.) Taking a similar approach as in \cite{Stephanov:2012ki,Stephanov:2014dma}, we start from the equation of motions for these fermions in their local rest frame
\begin{eqnarray} \label{eq_eom}
\sqrt{G_\pm}\ \dot{\vec{x}}=\frac{\vec{p}}{p}\pm\frac{\vec{\omega}}{p}  \quad , \quad \sqrt{G_\pm}\ \dot{\vec{p}}=2\vec{p}\times\vec{\omega},
\end{eqnarray}
where $\sqrt{G_\pm}=1\pm\vec{p}\cdot\vec{\omega}/p^2$ and $\omega=|\vec{\omega}|$ is the global rotational angular speed of the system. The corresponding kinetic equations can be written as:
\begin{eqnarray}
\partial_t f_{\pm}+\dot{\vec{x}}\cdot\partial_{\vec{x}}f_\pm+\dot{\vec{p}}\cdot\partial_{\vec{p}}f_\pm=C_\pm[f_+, f_-]  \,\, .
\end{eqnarray}
Integrating these equations and using Eq.~(\ref{eq_eom}), we obtain:
\begin{eqnarray} \label{eq_density}
&&\partial_t\int_{\vec{p}}\ \sqrt{G_\pm} f_\pm+\int_{\vec{p}}\ (\frac{\vec{p}}{p}\pm\frac{\vec{\omega}}{p})\cdot\partial_{\vec{x}}f_\pm   \nonumber\\
&& \qquad +\int_{\vec{p}}\ 2(\vec{p}\times\vec{\omega}) \cdot\partial_{\vec{p}}f_\pm
 =\int_{\vec{p}}\ \sqrt{G_\pm}C_\pm  \, . \quad
\end{eqnarray}
The last term of L.H.S is zero after integration by part. Also $\int_{\vec{p}}\equiv \int \frac{d^3\vec p}{(2\pi)^3}$.

We now examine small fluctuations in the net (vector) density on top of certain background equilibrium distribution $f_{0 \pm}(p;T,\mu_0)$. Similar to the analysis in~\cite{Stephanov:2014dma}, we parameterize the density fluctuations as:
\begin{eqnarray}
&&  f_\pm(t,\vec{x}, \vec{p}) = f_{0 \pm}(p) + \delta f_\pm(t,\vec{x}, \vec{p})    \,\, , \\
&&  \delta f_\pm =  \pm \left[\partial_p f_{0 \pm}(p) \right] \int d\nu d^3k e^{i(\nu t-\vec{k}\cdot\vec{x})}h(\nu, \vec{k}, \vec{p})  \,\, .  \,\,\,\,
\end{eqnarray}
where the $ \delta f_\pm$ have been expanded in Fourier modes. Subjecting the above into Eq.(\ref{eq_density}), and taking a difference to yield the time evolution of net density, we obtain the following relation in linear order of fluctuations,
\begin{eqnarray} \label{eq_mode}
&&\nu\int_{\vec{p}} \ \left[ \partial_p f_{0+}(p)+\partial_p f_{0-}(p)\right] h(\nu, \vec{k}, \vec{p})\nonumber\\
&&=\vec{k}\cdot\int_{\vec{p}} \ \frac{\vec{\omega}}{p} \left [ \partial_p f_{0+}(p)-\partial_p f_{0-}(p)\right ] h(\nu, \vec{k}, \vec{p})
\,\, , \,\,\,\,\,
\end{eqnarray}
where we have used the facts that (a) the equilibrium distribution $f_{0\pm}$ is a space-time independent fixed point of the collision kernel,  (b) $\int_{\vec{p}}\vec{p}\ z(p)=0$ for any  $z(p=|\vec p|)$, and (c) the collision terms from fluctuations vanish because of charge conservation constraint.

Let us then examine the low frequency, long wavelength limit, $\nu\to 0$ and $k \to 0$. The hydrodynamic zero mode in this limit arises from $\delta f_\pm \to \pm [\partial_p f_{0 \pm}(p)] H(t,\vec x)$ which implies $h$ becoming independent of $\vec p$.  This allows one to perform integrations over $\vec p$ in (\ref{eq_mode}), and obtain:
\begin{eqnarray}
\left[  \chi_{\mu_0}\ \nu -  C \ \vec{k}\cdot\vec{\omega} \right ]  h(\nu, \vec{k}) = 0,
\end{eqnarray}
where $\chi_{\mu_0}$ is the thermodynamic susceptibility defined in equilibrium, $\chi_{\mu_0} (T, \mu_0)=\partial n/\partial \mu|_{T,\mu_0}$ with net charge density $n=\frac{1}{(2\pi)^3}\int_{\vec{p}}\sqrt{G}(f_{0+}-f_{0-})$. The constant above is defined by $C=-\int_{\vec p} (1/p)\left [ \partial_p f_{0+}(p)-\partial_p f_{0-}(p)\right ] = \frac{1}{2\pi^2}\int dp [f_{0+}(p) -f_{0-}(p) ]$, and for the Fermi-Dirac distribution $C=\mu_0/(2\pi^2)$. This allows one to immediately identify a hydrodynamic collective excitation that propagates along the vorticity direction, $\vec k \parallel \vec \omega$, with the following dispersion relation:
\begin{eqnarray}
\nu = V_{\Omega} \, |\vec k| \quad, \quad  V_{\Omega}=\frac{\mu_0 \, \omega}{2\pi^2\chi_{\mu_0}}.
\end{eqnarray}
Notably, the so-obtained CVW speed $V_\Omega$ agrees exactly with that in Eq.(\ref{eq_cvw}).

{\it CVW-induced charge transport.---} We now discuss interesting charge transport phenomena induced by the CVW. To be concrete we consider different initial density fluctuations on top of a background medium with vorticity $\vec \omega=\omega \hat{x}$ and  CVW speed $V_{\Omega}$ given by (\ref{eq_cvw}).

Let us first consider purely axial charge density fluctuation in the initial condition, $F^A_i(x)$ at $t=0$ (we have suppressed ``trivial'' coordinates $y,z$). This can be cast into initial conditions for right-handed and left-handed density fluctuations, $F^{R/L}_i=\pm F^A_i / 2$. The subsequent evolution via the wave equations simply yields $(\delta n)^{R/L}_t=\pm F^A_i(x\mp V_\Omega t)/2$. We are interested in the transport of vector charge density which is an observable quantity. This can be obtained as follows:
\begin{eqnarray}
(\delta n)^V_t  &=&   \left[ F^A_i(x - V_\Omega t) - F^A_i(x +  V_\Omega t) \right] /2 \nonumber \\
&\approx& [-\partial_x F^A_i(x)] V_\Omega\, t
\end{eqnarray}
where the second line is true for small $V_\Omega t$. This implies a separation effect of vector charge along the vorticity direction: when initial axial charge fluctuation is positive and concentrated around $x=0$ (e.g. a Guassian form) then CVW will transport positive/negative vector charges toward $\pm \hat{x}$ directions respectively, leading to a {\it charge dipole moment} in parallel to $\vec \omega$ with strength $|d_\Omega|$ proportional to CVW speed $V_\Omega$ and propagation time; when the initial axial fluctuation is negative, the dipole moment flips. 

Let us then consider purely vector charge density fluctuation in the initial condition, $F^V_i(x)$ at $t=0$. Following similar procedure, the vector charge density from CVW evolution is given by:
\begin{eqnarray}
(\delta n)^V_t  &=&   \left[ F^V_i(x - V_\Omega t) +  F^V_i(x +  V_\Omega t) \right] /2 \nonumber \\
&\approx& F^V_i(x) + [\partial^2_x F^V_i(x)/2] (V_\Omega\, t)^2
\end{eqnarray}
where again the second line is true for small $V_\Omega t$. This transport process leads to a quadrupole moment of vector charge density along $\vec \omega$. Consider the initial fluctuation to be positive and concentrated around $x=0$ (e.g. a Guassian form), then $[\partial^2_x F^V_i(x)/2]$ is positive (negative) at large (small) $|x|$, implying concentration of positive charges away from $x=0$ toward both directions along $\vec \omega$. The resulting  quadrupole $|q_\Omega|$ is proportional to $(V_\Omega t)^2$.

{\it Experimental observable in heavy ion collisions.---} Our discussions on the CVW and its induced charge transport effects so far are rather general. We now consider its possible experimental manifestation in a concrete system, namely, rotating quark-gluon plasma created in off-central heavy ion collisions. The global rotation points in the out-of-plane direction. In such a QGP, CVW occurs for each light flavors, e.g. $u,d$ quarks and possibly $s$ quarks as well and transports flavor charges toward the two ``tips'' of the QGP fireball leading to a quadrupole charge distribution on the transverse plane. Here we make a first estimate of such effect.
\begin{figure}[!hbt]
\begin{center}
\includegraphics[width=0.3\textwidth]{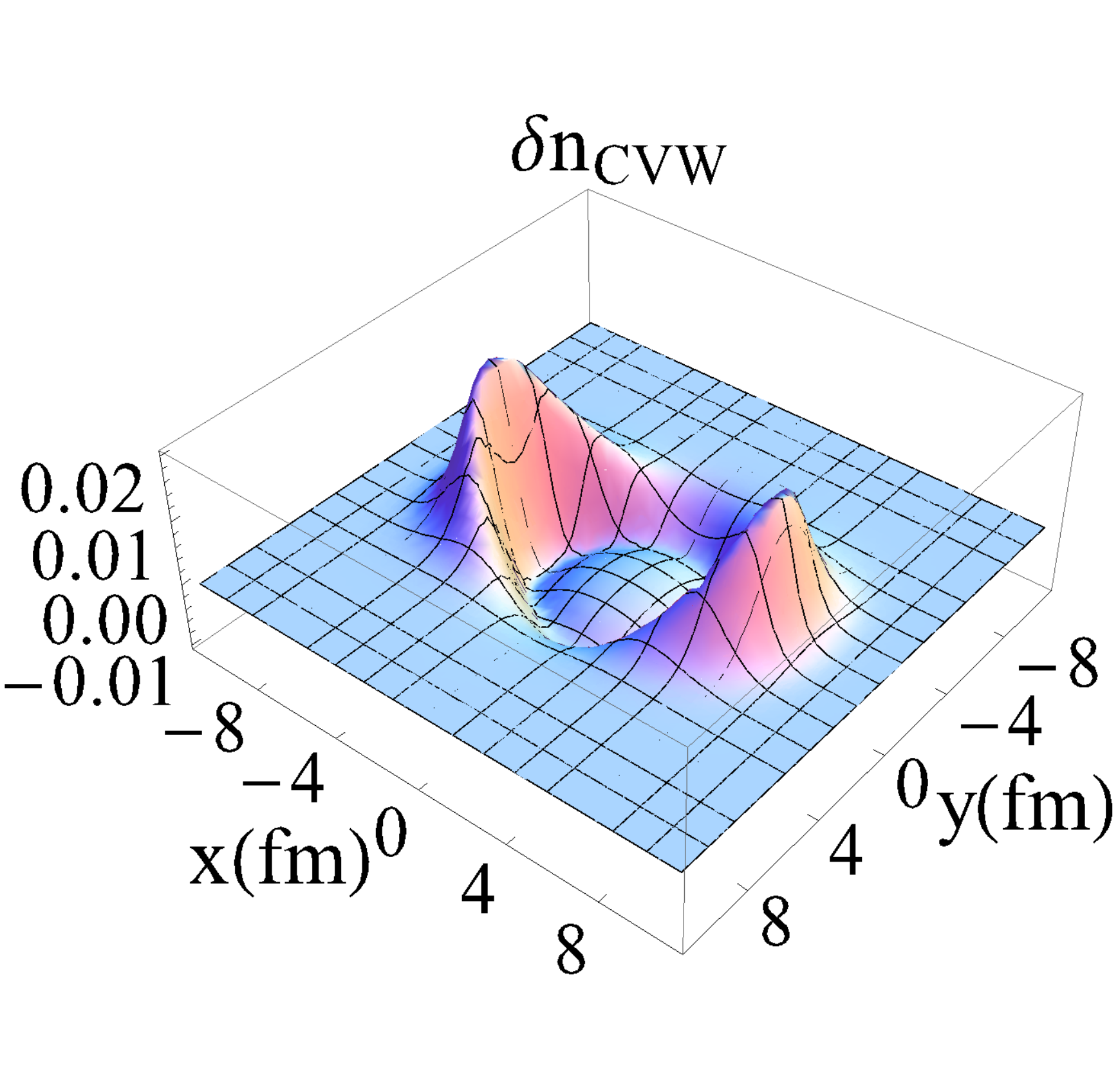} \vspace{-0.4in}
\caption{CVW-induced flavor charge density profile.} \vspace{-0.2in}
\label{fig_q}
\end{center}
\end{figure}

Let us first quantify the  quadrupole moment $q^f_\Omega$ resulting from CVW for a single quark flavor. We use the participant density from Glauber model as an initial condition for the flavor charge fluctuation and study the dependence of $q^f_\Omega$ on the key parameter $V_\Omega$ by solving the CVW equation. An illustration of the CVW-transported flavor charge density distribution at $\tau=8\rm fm$ (with beam energy $\sqrt{s}=200\rm GeV$,  impact parameter $b=7\rm fm$, initial time $\tau_0=0.6\rm fm$) is shown in Fig.\ref{fig_q}: a quadrupole pattern is evident. The quadrupole moment can be obtained by integrating the density distribution $q^f_\Omega= [\int dx dy (\delta n_f) \cos(2\phi_s)] / [\int dx dy (\delta n_f) ] $ . We have computed this quantity with the results:
$q^f_\Omega\simeq  -0.03 (V_{\Omega} \Delta\tau)^2$  with $\Delta\tau $ (in $fm/c$) the propagation time in QGP. The numerical coefficient is for minimum bias events and it varies at $\sim 15\%$ level across centrality. The minus sign is merely due to convention of defining azimuthal angle $\phi_s$ with respect to the in-plane direction.

Clearly we need a plausible estimate of $V_\Omega$.  Using the ``A Multi-Phase Transport''(AMPT) model~\cite{Lin:2004en}, our simulations suggest an initial value of $\vec \omega$ at about  $0.5 fm^{-1}$~\cite{vorticity,Becattini:2015ska,Becattini:2013vja,Florchinger:2011qf}, and lattice susceptibility at initial temperature $T_0\sim 350\rm MeV$  is about $\chi_f\sim 3  \rm fm^{-2} $~\cite{Borsanyi:2011sw,Bazavov:2013uja}, with both decreasing as QGP expands. Using a background density $\mu_0$ of $0.1\sim 1 \rm fm^{-1}$ one gets an estimate  $<V_\Omega> \simeq 10^{-2}\sim 10^{-3}$, leading to an induced quadrupole $|q^f_\Omega|$ at $\sim 10^{-4}$ level.
Such estimate is very sensitive to $\mu_0$. By going to lower beam energy   or by selecting events with large baryon asymmetry, the background $\mu_0$  could  be considerably increased thus magnifying $|q^f_\Omega|$.

\begin{figure}[!hbt]
\begin{center}
\includegraphics[width=0.37\textwidth]{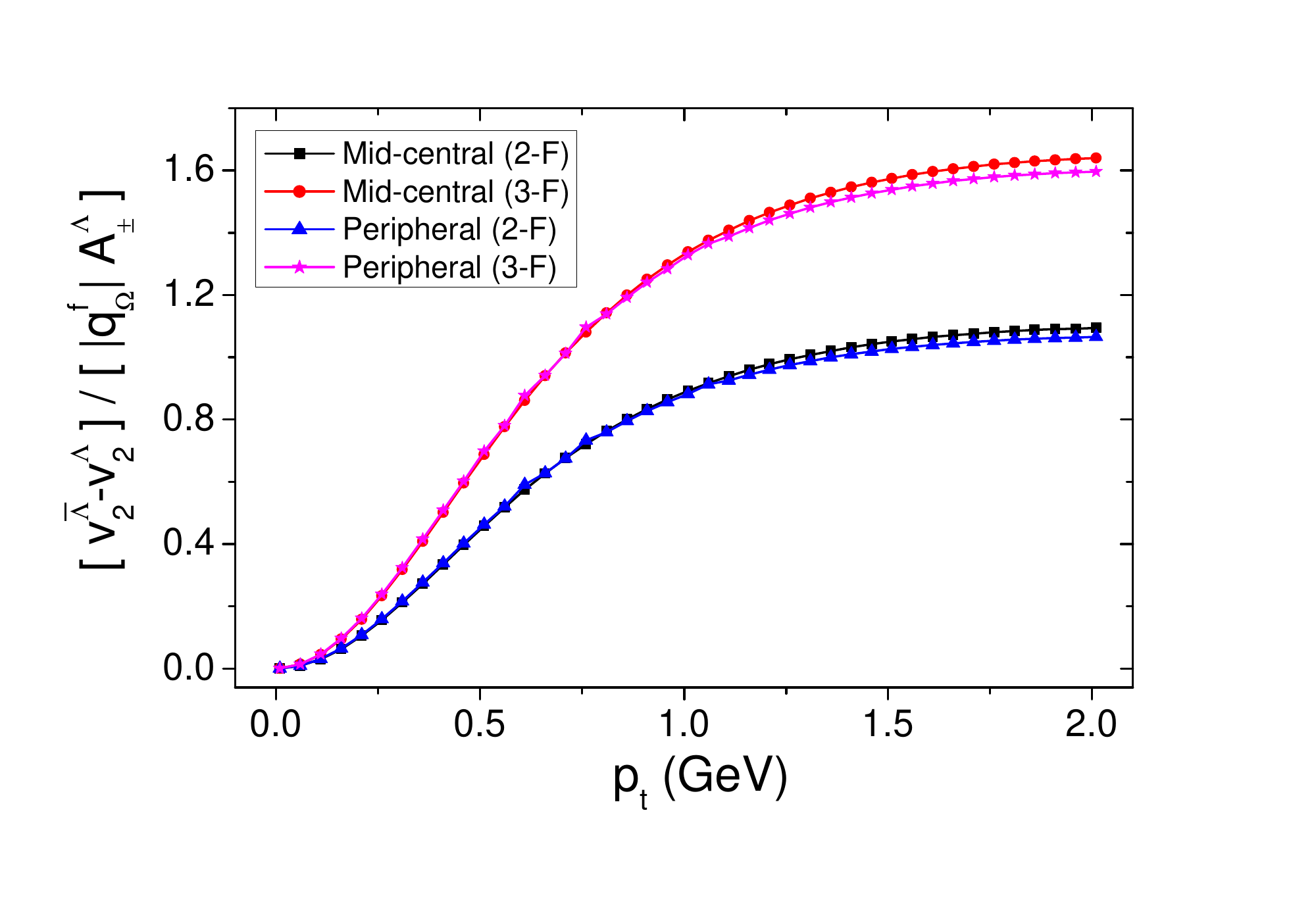}\vspace{-0.4in}
\caption{ Normalized $\bar{\Lambda}$ and $\Lambda$ elliptic flow splitting, $[v_2^{\bar{\Lambda}}-v_2^\Lambda]/[ |q^f_\Omega| A^\Lambda_{\pm}]$, for symmetric 2-flavor(2-F) and 3-flavor(3-F) cases. The mid-central and the peripheral correspond to for $15-30\%$ and $60-92\%$ centrality class (see~\cite{STAR_blastwave}).   } \vspace{-0.2in}
\label{fig_v2}
\end{center}
\end{figure}

 Flavor quadrupole implies that more baryons will be formed on the tips than on the equator of the fireball. The stronger in-plane radial flow will thus translate the quadrupole into baryon/anti-baryon $v_2$ splitting. This mechanism is in analogy to the electric charge quadrupole induced by CMW~\cite{Burnier:2011bf}. Suppose at the freeze-out, the flavor-wise chemical potential for quarks contains the CVW-induced quadrupole contribution $\delta \mu_f \propto 2 q^f_\Omega \cos(2\phi_s)$ (with $f=u,d,s$). The corresponding chemical potential for a given type of hadron can be determined from its constituent quark content, e.g. for $\Lambda$ baryon $\delta \mu_\Lambda \propto 2(q^u_\Omega+q^d_\Omega+q^s_\Omega) \cos(2\phi_s)$.  We particularly propose to use $\Lambda$ baryon which is electric charge neutral thus unaffected by possible CMW effect. We then use the STAR blast-wave model~\cite{STAR_blastwave} to compute the resulting differential flow splitting. As it is unclear how much the $s$ quark mass may reduce their chiral effects, we consider two extreme cases: a symmetric two-flavor (2-F) case $q^u_\Omega=q^d_\Omega=q^f_\Omega$ with $q^s_\Omega=0$, or a symmetric three-flavor (3-F) case $q^u_\Omega=q^d_\Omega=q^s_\Omega=q^f_\Omega$. From Cooper-Frye scheme it is easy to see $\Delta v_2=v_2^{\bar{\Lambda}}-v_2^\Lambda \propto  |q^f_\Omega| A^\Lambda_{\pm}$ with $A^\Lambda_{\pm}=(N^\Lambda-N^{\bar{\Lambda}})/(N^\Lambda+N^{\bar{\Lambda}})$ the $\Lambda$-asymmetry that is directly related to background density $\mu_0$  (in analogy to a similar relation in CMW case~\cite{Burnier:2011bf}).  The results for normalized flow splitting $\Delta v_2/[ |q^f_\Omega| A^\Lambda_{\pm}]$  are shown in Fig.\ref{fig_v2}. Note that while the curves for the two centralities appear close, they have rather different normalization as the $|q^f_\Omega|\sim V_\Omega^2$ strongly depends on centrality. Note also that the CVW predicts a particular slope for $\Delta v_2\propto A^\Lambda_{\pm}$ but may not exclude a finite intercept at $A^\Lambda_{\pm}=0$ with either sign. Needless to say, these are crude estimates and a realistic hydrodynamic modeling of CVW (that accounts for factors like time-dependent vorticity, susceptibility and diffusion) will be done in a future work. Given that, our results suggest that a CVW-induced signal could be detected and may give indications on chiral effects of strange flavor. Experimental measurements of $\Lambda$ and $\bar\Lambda$ $v_2$ are feasible (see e.g.~\cite{Abelev:2007qg}), and the predicted flow splitting may be measured for events binned according to their baryonic number asymmetry~\cite{exp}.

{\it Summary.---}  In summary, we have found a new gapless collective excitation in a rotating fluid system with chiral fermions, named as the Chiral Vortical Wave. We derive the wave equation for CVW and determine its speed from both hydrodynamic and chiral kinetic theory. We demonstrate that the CVW can induce flavor quadrupole in rotating quark-gluon plasma in heavy-ion collisions which in turn split the elliptic flow for $\Lambda$ baryons. Such proposal could be tested with  future experimental data.

As a final remark, while the proposed CVW bears certain similarity to the CMW, it is a completely new phenomenon that provides an independent way of manifesting chiral anomaly.  In the context of heavy ion collisions, the vorticity lasts significantly longer than the lifetime of strong magnetic field and may induce more robust signal. In certain spin-orbit coupled cold Fermi gases that could simulate chiral anomaly effects~\cite{cold_atom}, only the vorticity driven effects can be easily and directly induced with those charge neutral atoms. With such unique merit the CVW has its own significance and interest for the study of anomalous effects in a wide range of physical systems.

\vskip0.2cm

{\it Acknowledgments.---} We thank Shu Lin and Yi Yin for  discussions. JL is grateful to Aihong Tang and Zhangbu Xu for very helpful discussions on experimental measurements. The research of YJ and JL is supported by National Science Foundation (Grant No. PHY-1352368). The research of XGH is supported by Fudan University (Grant No. EZH1512519) and Shanghai Natural Science Foundation (Grant No. 14ZR1403000). JL also thanks the RIKEN BNL Research Center for partial support.

 \vfil

\end{document}